\documentclass[a4,11pt]{article}
\usepackage{graphics}
\usepackage[dvips]{graphicx}

\usepackage{bm} 
\usepackage{amsmath}
\usepackage{amssymb}

\title{Effect of transition-metal elements on the electronic 
properties of quasicrystals and complex aluminides}

\author{
G. Trambly de Laissardi\`ere$^1$ and D. Mayou$^2$ \\ 
{\it $^1$\,Laboratoire de Physique Th\'eorique et Mod\'elisation,\\
CNRS and}
{\it Universit\'e de Cergy--Pontoise,  95302 Cergy--Pontoise, France}\\
{\it $^2$\,Institut N\'eel, CNRS and Universit\'e Joseph Fourier, B\^at D,}\\
{\it B.P. 166, 38042 Grenoble Cedex 9, France}
}
\date{}

\begin{document}

\newcommand{\spd}{$sp$--$d$ }
\newcommand{\ef}{E_{\rm F}}
\newcommand{\Ang}{{\rm \AA}}

\newcommand{\be}{\begin{equation}}
\newcommand{\ee}{\end{equation}}
\newcommand{\ben}{\begin{eqnarray}}
\newcommand{\een}{\end{eqnarray}}
\newcommand{\beq}{\begin{equation}}
\newcommand{\eeq}{\end{equation}}
\newcommand{\B}{\mathrm{B}}
\newcommand{\NB}{\mathrm{NB}}
\newcommand{\RRe}{\mathrm{Re}}
\newcommand{\IIm}{\mathrm{Im}}

\begin{flushright}

\includegraphics[width=7.5cm]{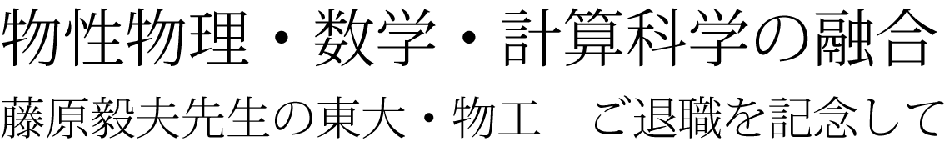}

{\small \it
Publication in the honor of Prof. T. Fujiwara.

\vskip -.2cm
Editors: Y. Hatsugai, M. Arai, S. Yamamoto, 2007, p. {128-144}
}
\end{flushright}

\thispagestyle{empty}

\vskip .5cm
\begin{center}
{\LARGE \bf Effect of transition-metal elements on the electronic properties of quasicrystals and complex aluminides}

\vspace{.3cm}
{\Large Guy T{\footnotesize RAMBLY} de L{\footnotesize AISSARDI\`ERE}\,$^1$, 
Didier M{\footnotesize AYOU}\,$^2$}

\vspace{.2cm}
{\it $^1$\,Laboratoire de Physique Th\'eorique et Mod\'elisation,
CNRS et}\\ {\it Universit\'e de Cergy--Pontoise, France.}\\
{\it guy.trambly@u-cergy.fr}\\
{\it $^2$\,Institut N\'eel, CNRS et Universit\'e Joseph Fourier,}
{\it Grenoble, France.}\\
{\it didier.mayou@grenoble.cnrs.fr}
\end{center}

\section{Introduction}

%
%

It is with great pleasure that we contribute to this book in  honor of 
Prof. Takeo Fujiwara. 
GTL enjoyed eighteen months of Prof.  
Fujiwara's hospitality at the University of Tokyo during the early 1990's. 
At that time the work of Prof. Fujiwara in the field of electronic   
structure of quasicrystals had already made a major contribution to the 
literature (see for instance \cite{FujiwaraSofArt}).
Since that  time our research owes much to his work.

Prof.  Fujiwara was the first who performed realistic  calculations 
of the electronic structure in quasicrystalline materials
without adjustable parameters (ab-initio calculations) \cite{Fujiwara89}. 
Indeed these complex  alloys \cite{Shechtman84} have very exotic
physical properties 
(see Refs. \cite{Berger94,Grenet00_Aussois} and Refs therein), 
and it
rapidly appeared  that realistic calculations on the 
actual quasicrystalline materials 
are necessary to understand the physical mechanism that govern this 
properties. 
In particular, these calculations allow to analyze numerically
the role of transition-metal elements which is essential in those materials.

In this paper, we briefly present our work on the role of transition-metal 
element in electronic structure and transport properties of quasicrystals 
and related complex phases. 
Several Parts of these works have been done 
or initiated
in collaboration with Prof. T. Fujiwara.

\section{Electronic structure}

\subsection{Ab-initio determination of the density of states}

\begin{figure}[t]
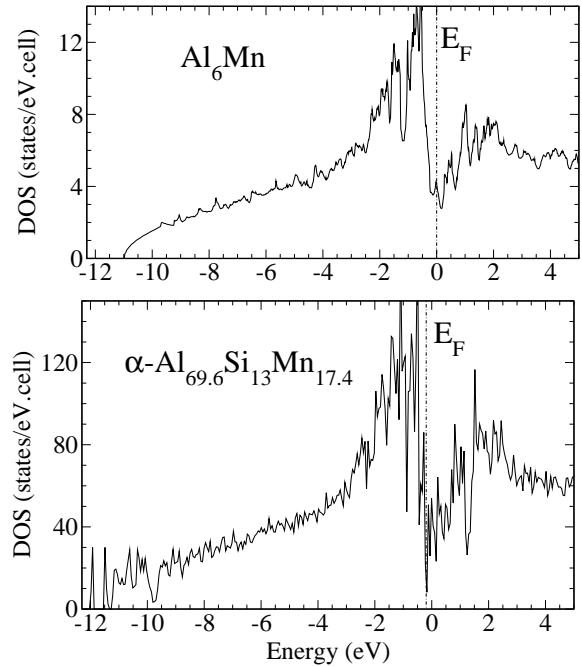

\begin{center}
\includegraphics[width=7.5cm]{al6mn_DOS_arttransp.eps}

\vskip 0.2cm
\includegraphics[width=7.5cm]{sugi_dAlSi_Zijl_b_6_all.eps}
\vspace{-0.4cm}
\caption{Ab-initio total DOS of $\rm Al_6Mn$ (simple crystal)
and $\alpha$-Al$_{69.6}$Si$_{13.0}$Mn$_{17.4}$ 
(approximant of  icosahedral quasicrystals) \cite{PMS05,Zijlstra03}.
\label{Fig_DOS_Al6Mn_sugi}}
\end{center}
\end{figure}

A way to study the electronic structure of quasicrystal is to 
consider the case of approximants. 
Approximants are
crystallines phases, with 
very large unit cell,
which reproduce the atomic order of quasicrystals locally.
Experiments indicate that approximant phases, like 
$\alpha$-AlMnSi, $\alpha$-AlCuFeSi, $R$-AlCuFe, etc., have transport
properties similar to those of quasicrystals \cite{Berger94,Quivy96}.
In 1989 and 1991, Prof.  Fujiwara performed the first 
numerical calculations of 
the electronic structure in realistic approximants
of quasicrystals \cite{Fujiwara89,Fujiwara91,Fujiwara93}.
He showed that their density of states 
(DOS, see figure \ref{Fig_DOS_Al6Mn_sugi}) is characterized
by  a depletion near the Fermi energy $\ef$,
called {\it ``pseudo-gap''},
in agreement with experimental results
(for review see Ref. \cite{Berger94,Mizutani02,Belin04})
and a Hume-Rothery stabilization \cite{Massalski78,PMS05}. 
The electronic structure of simpler crystals such as
orthorhombic $\rm Al_6Mn$,
cubic $\rm Al_{12}Mn$, present also a pseudo-gap near $\ef$
which is less pronounced than in complex approximants phases
(figure  \ref{Fig_DOS_Al6Mn_sugi}) \cite{PMS05}.

\subsection{Models to analyze the role of transition-metal element}

\subsubsection*{\spd hybridization model}

The role of the transition-metal (TM, TM = Ti, Cr, Mn, Fe, Co, Ni) elements 
in the pseudo-gap formation
has  been shown from experiments, 
ab-initio calculations 
and model analysis 
[4,13--19,11].
Indeed the formation of the pseudo-gap  results from
a strong \spd coupling associated to an ordered sub-lattice
of TM atoms \cite{Guy04_ICQ8,PMS05}.
Consequently, 
the electronic structure,
the magnetic properties and the stability, 
depend strongly on the TM positions,
as was shown from 
ab-initio calculations 
[28--33,20,21].

\subsubsection*{How an effective TM--TM interaction induces stability?}

\begin{figure}[t!]
\begin{center} 
\includegraphics[width=8cm]{PotMn_Mn_m0_c.eps} 
\vspace{-0.4cm}
\caption{Effective 
medium-range Mn--Mn interaction 
between two non-magnetic manganese atoms in a
free electron matrix which models aluminum atoms. 
\cite{PMS05}} 
\label{PotMn_Mn_m0} 
\end{center} 
\vspace{0.3cm}
\begin{center}
\includegraphics[width=8cm]{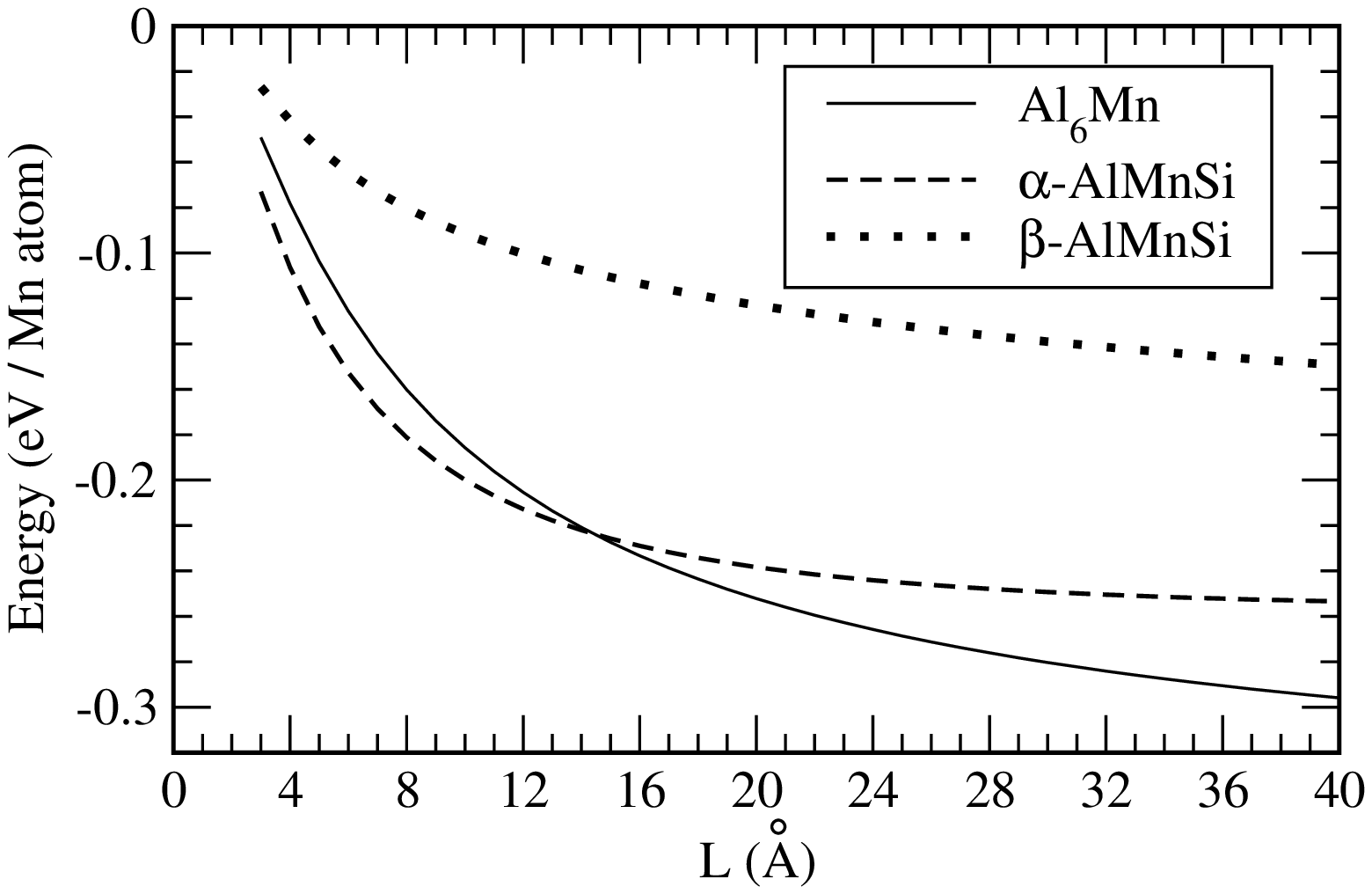} 
\vspace{-0.4cm}
\caption{Variation of the band energy
due to the effective Mn--Mn interaction
in   $\rm o$-$\rm Al_6Mn$,
$\rm \alpha$-AlMnSi and $\rm \beta$-$\rm Al_9Mn_3Si$.
\cite{Guy03}
}
\label{FigEBetaPhiAlphaAl6Mn}
\end{center}
\end{figure}

Just as for Hume-Rothery phases a description of the band energy
can be made in terms of pair interactions 
(figure \ref{PotMn_Mn_m0}) \cite{ZouPRL93,Guy04_ICQ8}.
Indeed, it has been shown that
an effective medium-range Mn--Mn interaction mediated by the
$sp$(Al)--$d$(Mn) hybridization plays a determinant role
in the occurrence of the pseudo-gap \cite{Guy04_ICQ8}.
We have shown that this interaction,  
up to distances 
10--20\,$\rm \AA$,
is essential in stabilizing these phases, 
since it can create a Hume-Rothery pseudo-gap  close to 
$\ef$. 
The band energy is then minimized as shown 
on figure \ref{FigEBetaPhiAlphaAl6Mn} \cite{Guy03,PMS05}.

The effect of these 
effective Mn--Mn interactions has been also studied 
by several groups
\cite{ZouPRL93,Guy03,GuyPRL00} (see also Refs in \cite{PMS05}). 
It has also explained the origin of  large  
vacancies in the hexagonal $\beta$-$\rm Al_9Mn_3Si$ and  
$\varphi$-$\rm Al_{10}Mn_3$ phases on some sites, whereas  
equivalent sites are occupied by Mn in  
$\rm \mu$-$\rm Al_{4.12}Mn$ 
and $\rm \lambda$-$\rm Al_4Mn$, and by Co 
in $\rm Al_5Co_2$ \cite{Guy03}. 
On the other hand, an
spin-polarized effective Mn--Mn  
interaction is also determinant for the  
existence (or not) of magnetic moments in AlMn 
quasicrystals and approximants \cite{GuyPRL00,Virginie2,Duc03}. 

The analysis can be applied to
any Al(rich)-Mn phases, where a small number of 
Mn atoms are embedded in the free electron 
like Al matrix. 
The studied effects 
are not specific to quasicrystals 
and their approximants, but they are more important
for those alloys.
Such a
Hume-Rothery stabilization, governed
by the effective 
medium-range Mn--Mn interaction, 
might therefore be intrinsically linked 
to the emergence of quasi-periodicity in Al(rich)-Mn 
system.

\subsubsection*{Cluster Virtual Bound states}

One of the main results of the ab-initio calculations performed by
Prof. Fujiwara for realistic approximant phases, is 
the small energy dispersion of electrons in the reciprocal space. 
Consequently,
the density of states of approximants is characterized by
{\it ``spiky''} peaks \cite{Fujiwara89,Fujiwara91,Fujiwara93,GuyPRB94_AlCuFe}.
In order to analyze the origin of this spiky structure of the DOS, we developed
a model that show  a new kind of localization by atomic cluster 
\cite{GuyPRB97}.

\begin{figure}[t]
\begin{center}
\includegraphics[width=7cm]{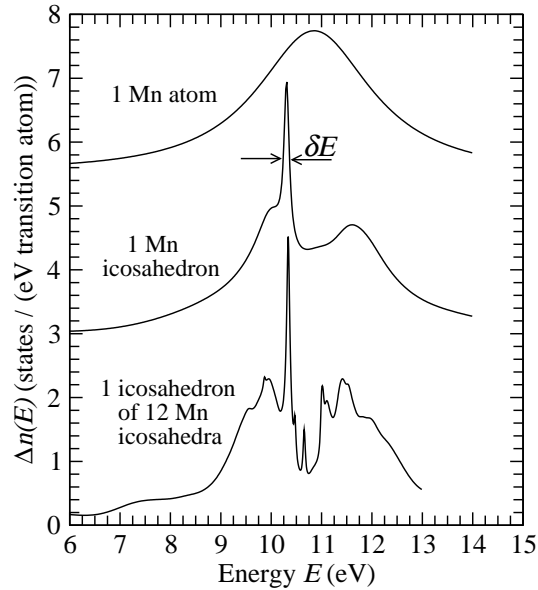}
\end{center}
\vspace{-0.4cm}
\caption{\label{DOS_Clusters}Variation $\Delta n(E)$ of the DOS due 
to Mn atoms. 
Mn atoms are embedded in a metallic medium (Al matrix).
From \cite{GuyPRB97}.}
\end{figure}

As for the local atomic order, 
one of the characteristics of the quasicrystals and approximants 
is the occurrence of atomic clusters on a scale of 10--30 $\rm \AA$ 
\cite{Gratias00}.  
The role of clusters has been much debated in particular by 
C. Janot \cite{Janot94}
and G. Trambly de Laissardi\`ere \cite{GuyPRB97}. 
Our model   
is based on a standard description of inter-metallic alloys. 
Considering the cluster embedded in a metallic medium,  
the variation $\Delta n(E)$ 
of the DOS due to the cluster is calculated.
For electrons, which have energy in the vicinity of the Fermi level, 
transition atoms (such as Mn and Fe) are strong scatters whereas 
Al atoms are weak scatters. 
In the figure \ref{DOS_Clusters} the variation, $\Delta n(E)$,
of the density of states  due to different clusters
 are shown. The Mn icosahedron is the actual Mn icosahedron  of the 
$\alpha$-AlMnSi approximant.
As an example of a larger cluster, we consider one icosahedron of Mn icosahedra.

$\Delta n(E)$ of clusters exhibits strong deviations from the 
Virtual Bound States (1 Mn atom)
\cite{Friedel56}.
Indeed several peaks and shoulders appear. 
The width of the most narrow peaks 
($50 - 100$\,meV)
are comparable to the fine peaks of the calculated DOS in the approximants
(figure \ref{Fig_DOS_Al6Mn_sugi}). 
Each peak indicates a resonance due to the scattering by the cluster. 
These peaks correspond to states {\it ``localized''}  
by the icosahedron or the icosahedron of icosahedra. 
They are not eigenstate, they have finite lifetime of the order of 
$\hbar / \delta E$, 
where $\delta E$ is the width of the peak. 
Therefore, the stronger the effect of the localization by cluster is, 
the narrower is the peak. 
A large lifetime is the proof of a localization, 
but in the real space these states have a quite large extension on 
length scale of the cluster. 

The physical origin of these states can be understood as follows. 
Electrons are scattered by the Mn atoms of a  cluster.
By an effect similar to that of a Faraday cage, 
electrons can by confined 
by the cluster provided that their wavelength $\lambda$ satisfies 
$\lambda \gtrsim l$, 
where $l$ is the distance between two Mn spheres. 
Consequently, we expect to observe such a confinement by the cluster.
This effect is a multiple scattering effect, and it is not due to an overlap 
between $d$-orbitals because Mn atoms are not first neighbor.

\section{Transport properties}

Quasicrystals have many fascinating electronic properties,
and in particular quasicrystals with high structural quality,
such as the icosahedral AlCuFe and AlPdMn alloys,
have unconventional conduction properties when compared with
standard inter-metallic alloys. Their conductivities can be as
low as 150--200\,($\rm \Omega\,cm)^{-1}$ 
(see Refs. \cite{Berger94,Grenet00_Aussois,Mayou93} and Refs. therein).
Furthermore the conductivity increases
with disorder and with temperature, a behavior just at the
opposite of that of standard metal. In a sense the most striking
property is the so-called {\it ``inverse Mathiessen rule''}
according to which the {\it increases of conductivity} due to
different sources of disorder seems to be {\it additive}.
This is just the opposite that happens with normal metals
where the increases of resistivity due to several sources
of scattering are {\it additive}.
An important result is also that many approximants of
these quasicrystalline phases have similar conduction properties.
For example the crystalline $\alpha$-AlMnSi phase with a unit cell
size of about 12\,$\rm \AA$ and 138 atoms in the unit cell
has a conductivity of about 300\,($\rm \Omega\,cm)^{-1}$
at low temperature \cite{Berger94}.

\begin{figure}[t!]
\begin{center}
\includegraphics[width=10cm]{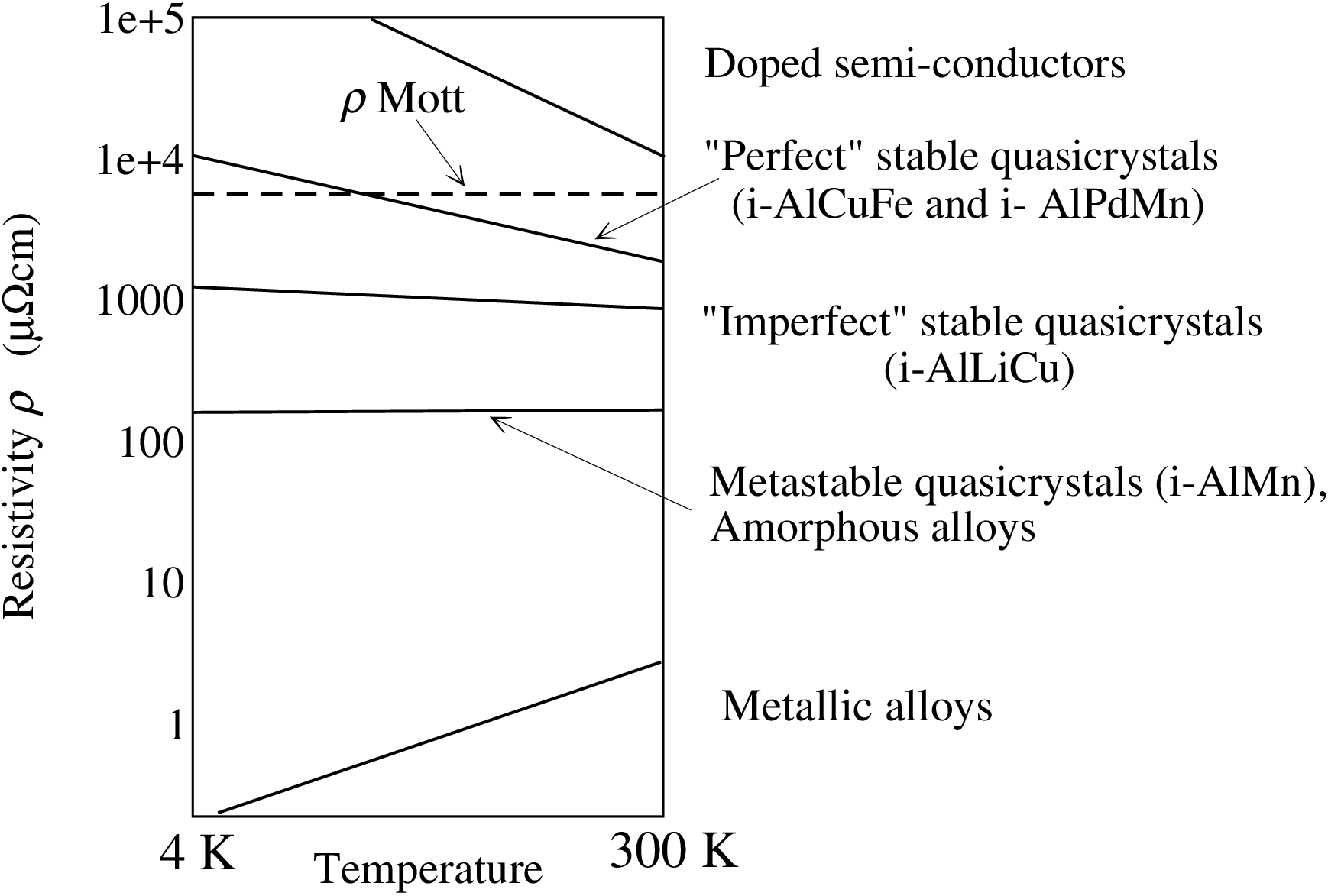}
\vspace{-.4cm}
\caption{Schematic temperature dependencies of 
the experimental resistivity of quasicrystals,  amorphous 
and metallic crystals.
\label{Fig_Resistivite_schema}}
\end{center}
\vspace{.1cm}
\begin{center}
\begin{tabular}{cc}
\begin{minipage}[c]{6.4cm}
\includegraphics[width=6.3cm]{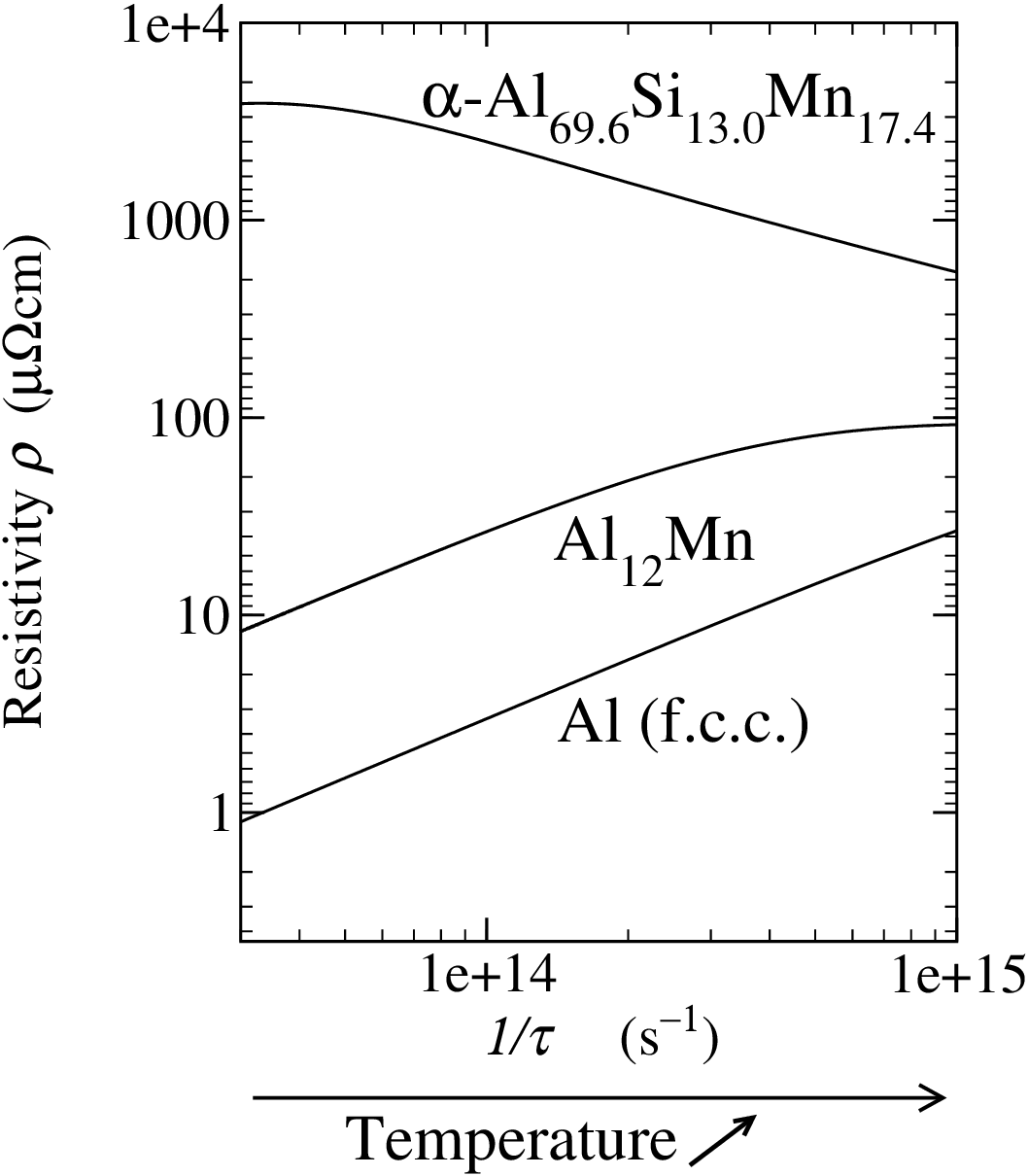}
\end{minipage}
&
\begin{minipage}[c]{4.5cm}
\caption{\label{Fig_ResistiviteAbInitio} Ab-initio electrical resistivity 
versus inverse scattering time,
in cubic approximant $\alpha$-Al$_{69.6}$Si$_{13.0}$Mn$_{17.4}$,
pure Al (f.c.c.), and cubic Al$_{12}$Mn.}
\end{minipage}
\end{tabular}
\end{center}
\end{figure}

\subsection{Small Boltzmann velocity}

Prof.  Fujiwara {\it et al.} was the first to show  that the electronic
structure of AlTM approximants and related phases is
characterized by two energy scales
\cite{Fujiwara89,Fujiwara91,Fujiwara93,GuyPRB94_AlCuFe,GuyPRBAlCuCo}
(see previous section).
The largest energy scale, of about $0.5 - 1$\,eV, 
is the width of the pseudogap
near the Fermi energy $E_{\rm F}$. 
It
is related to the Hume--Rothery stabilization via the scattering of
electrons by the TM sub-lattice because of a strong \spd
hybridization. 
The smallest energy scale, less
than 0.1\,eV, is characteristic of the small dispersion of the
band energy $\rm E({\bf k})$. 
This energy scale seems more
specific to phases related to the  quasi-periodicity.  
The first consequence on transport
is a small
velocity at Fermi energy, Boltzmann velocity, 
$V_{\B} =(\partial E/ \partial
k)_{E=E_{\rm F}}$. From numerical calculations, 
Prof.  Fujiwara {\it et al.}
evaluated the Bloch--Boltzmann dc conductivity $\sigma_{\B}$ in the
relaxation time approximation. With a realistic value of
scattering time, $\tau \sim 10^{-14}$\,s~\cite{Mayou93}, one
obtains $\sigma_{\B} \sim 10-150\,({\rm \Omega cm})^{-1}$ for a
$\alpha$-AlMn model~\cite{Fujiwara93} and 1/1-AlFeCu
model~\cite{GuyPRB94_AlCuFe}. This  corresponds to the measured
values~\cite{Berger94,Quivy96}, which are anomalously low for
metallic alloys. For
decagonal approximant the anisotropy found experimentally in the
conductivity is also reproduced correctly~\cite{GuyPRBAlCuCo}.

\subsection{Quantum transport in Quasicrystals and approximants}

The semi-classical Bloch--Boltzmann description of transport gives
interesting results for the intra-band conductivity in crystalline
approximants, but it is insufficient to take into account many aspects
due to the special localization of electrons by
the quasi-periodicity 
(see Refs. 
[34--43]
and Refs. therein). 
Some specific transport
mechanisms like the temperature dependence of the conductivity
(inverse Mathiessen rule, the defects influence,
the proximity of  a metal\,/\,insulator transition), 
require to go
beyond a Bloch--Boltzmann analysis. 
Thus, it appears that in quasicrystals and related complex metallic 
alloys a new type of breakdown of the semi-classical Bloch-Boltzmann 
theory operates.
In the literature,
two different unconventional transport
mechanisms have been proposed for these materials.
Transport could be dominated, for short relaxation time $\tau$ by
hopping between ``{\it critical} localized states'', whereas for
long time $\tau$ the regime could be dominated by non-ballistic
propagation of wave packets between two scattering events.

We develop a theory of quantum transport that applies to a normal 
ballistic law but also to  these specific diffusion laws. 
As we show phenomenological models based on this theory describe correctly 
the experimental transport properties 
\cite{MayouPRL00,PRL06,MayouRevueTransp}
(compare figures \ref{Fig_Resistivite_schema} 
and \ref{Fig_ResistiviteAbInitio}).

\subsection{Ab-initio calculations of quantum transport}

\begin{figure}[t!]
\begin{center}
\includegraphics[width=8cm]{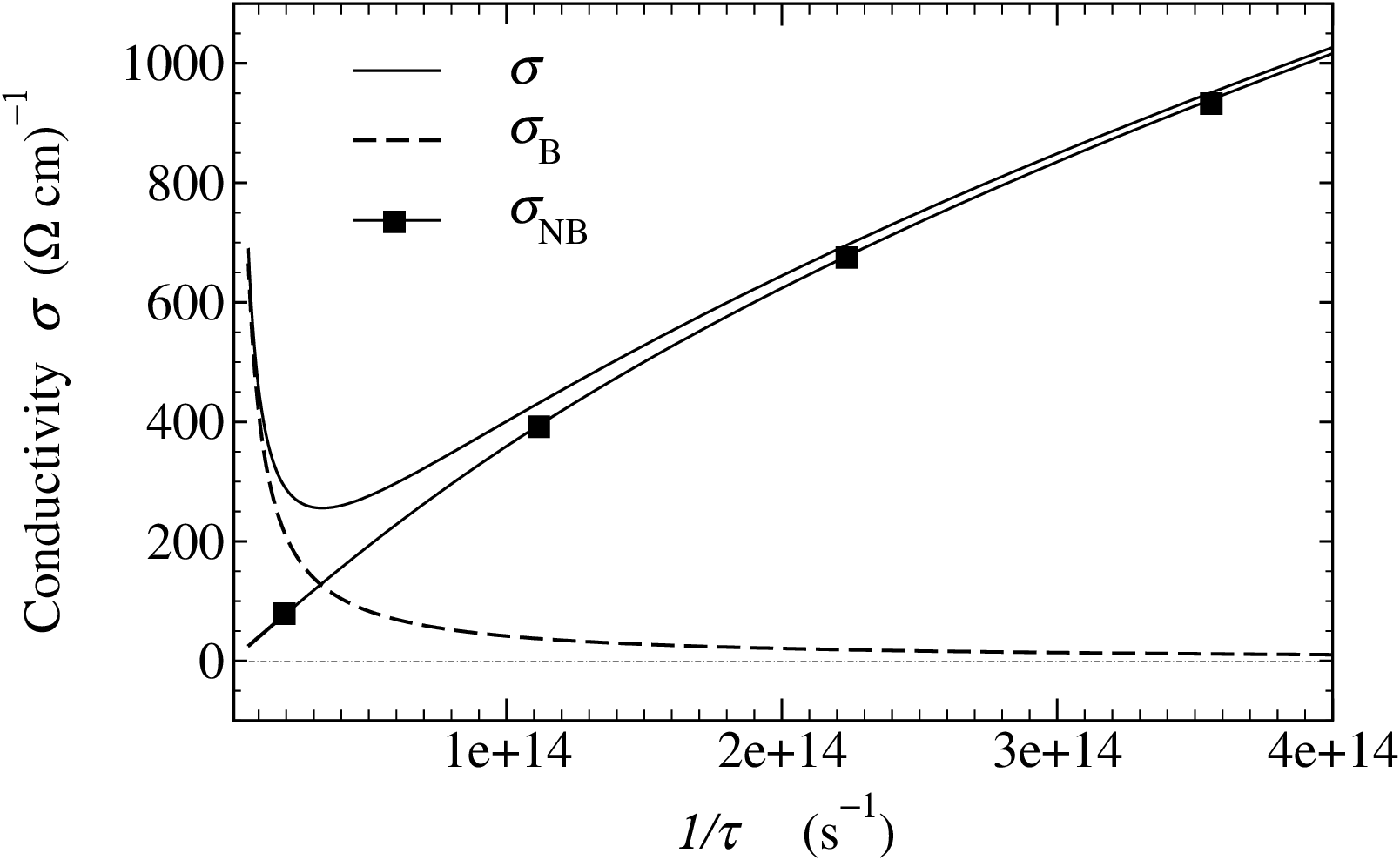}
\end{center}
\vspace{-0.5cm}
\caption{ \label{Fig_Conductivity_t6_T_e.012D.001}
Ab-initio dc-conductivity $\sigma$ in cubic approximant
$\alpha$-Al$_{69.6}$Si$_{13.0}$Mn$_{17.4}$ versus
inverse scattering time. \cite{PRL06}
}
\vspace{0.3cm}
\begin{center}
\includegraphics[width=8cm]{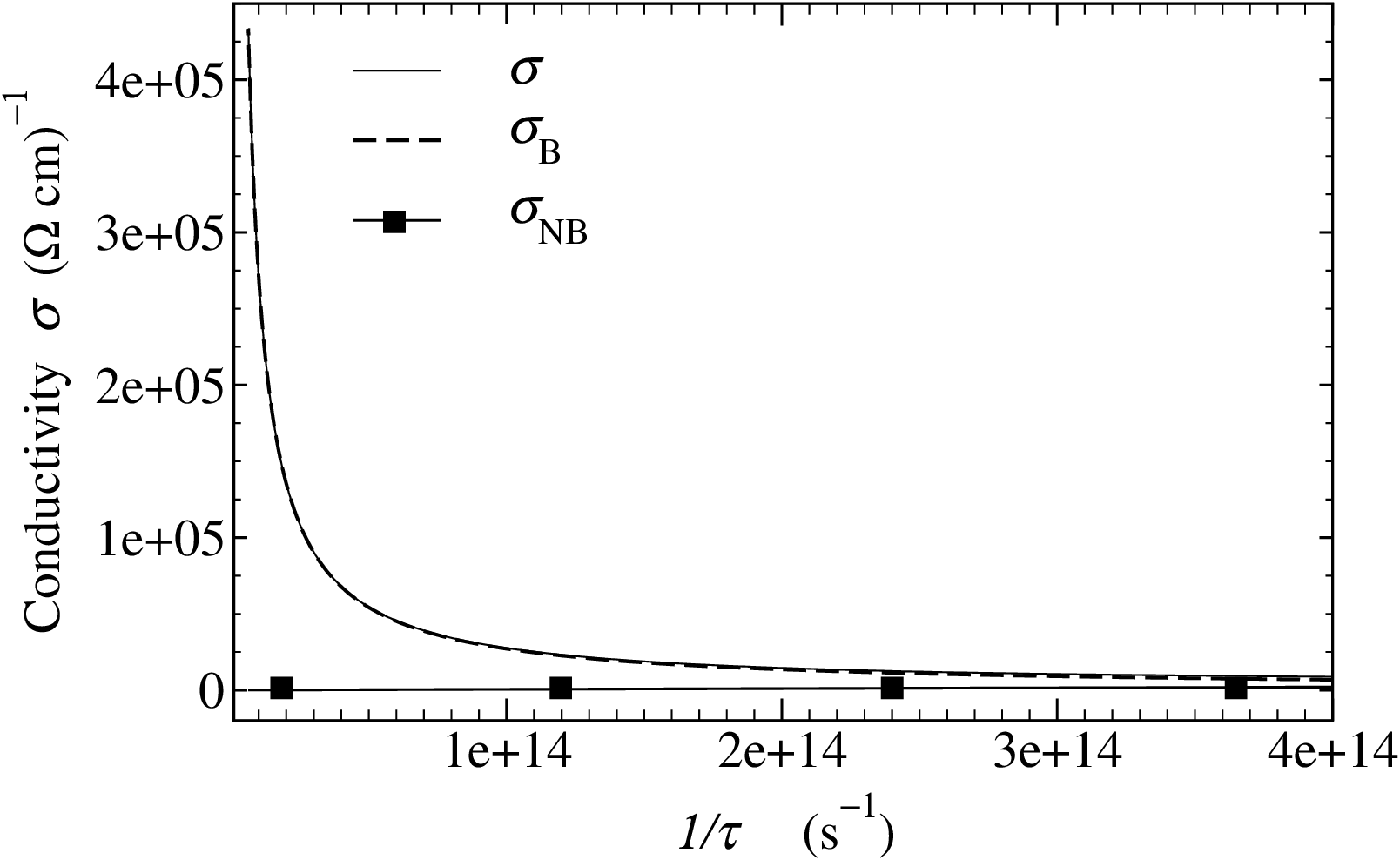}
\end{center}
\vspace{-0.5cm}
\caption{\label{Fig_Sig_aAlCuSi} 
Ab-initio dc-conductivity $\sigma$ in an hypothetical cubic approximant
$\alpha$-Al$_{69.6}$Si$_{13.0}$Cu$_{17.4}$ versus
inverse scattering time. \cite{MayouRevueTransp}
}
\end{figure}

According to the Einstein relation the conductivity
$\sigma$ depends on the diffusivity $D(E)$ of electrons 
of energy $E$
and the density of states $n(E)$ (summing the 
spin up and spin down contribution).
We assume that $n(E)$
and $D(E)$ vary weakly on the thermal energy scale $kT$,
which is justified here.
In that case, the Einstein formula
writes
\ben
\sigma = e^2 n(E_{\mathrm{F}})D(E_{\mathrm{F}})
\een
where $E_{\mathrm{F}}$ is the chemical potential
and $e$ is the electronic charge.
The temperature dependence of $\sigma$ is 
due to the variation of the diffusivity $D(E_F)$ with
temperature.
The central quantity is thus the diffusivity which 
is related to quantum diffusion.
Within the relaxation time approximation, the diffusivity 
is written \cite{MayouPRL00}
\ben
D(E) =\frac{1}{2} \int_{0}^{+\infty} C_0(E,t) 
\,{\rm e}^{-|t| / \tau}\,{\rm d} t
\label{DRTA}
\een
where 
$C_0(E,t)=\Big\langle V_x(t)V_x(0) + V_x(0)V_x(t) \Big\rangle_E$
it the velocity correlation functions  without disorder, 
and $\tau$ is the relaxation time.
Here, the effect of defects and temperature  
(scattering by phonons\,...)
is taken into account through the relaxation time 
$\tau$. $\tau$ decreases as disorder increases.
In the case of crystals phases (such as approximants of quasicrystals),
one obtains \cite{PRL06,MayouRevueTransp}:
\begin{eqnarray}
\sigma  &=& \sigma_{\B} ~+~ \sigma_{\NB} 
\label{Eq_sigma0}\\
\sigma_{\B}  = e^2 n(E_{\mathrm{F}})\,V_{\B}^2 \,\tau 
&{\rm and}& \sigma_{\NB} = e^2 n(E_{\mathrm{F}})\,\frac{L^2(\tau)}{\tau}
\label{Eq_sigma}
\end{eqnarray}
where $\sigma_{\B}$ is actual the Bolzmann contribution 
to the conductivity
and $\sigma_{\NB}$ a non-Boltzmann contribution.
$L^2(\tau)$ is smaller than the square of the unit cell size $L_0$. 
$L^2(\tau)$ can be calculated numerically for the ab-initio 
electronic structure \cite{PRL06}.
From (\ref{Eq_sigma0}) and (\ref{Eq_sigma}), 
it is clear that the Bolzmann term dominates 
when $L_0 \ll V_{\B}\tau$:
The diffusion of electrons is then ballistic, 
which is the case in normal metallic crystals.
But, when $L_0 \simeq V_{\B}\tau$, i.e. when the Bolzmann velocity 
$V_{\B}$ is very low, the non-Bolzmann term is essential.
In the case  of $\alpha$-Al$_{69.6}$Si$_{13.0}$Mn$_{17.4}$
approximant (figure \ref{Fig_Conductivity_t6_T_e.012D.001}) \cite{PRL06}, 
with realistic value of $\tau$ 
($\tau$ equals a few $10^{-14}$\,s \cite{Mayou93}),
$\sigma_{\NB}$ dominates and
$\sigma$ increases when $1/\tau$ increases, i.e. when
defects or temperature increases, in agreement with 
experimental measurement 
(compare  figures 
\ref{Fig_Resistivite_schema} 
and \ref{Fig_ResistiviteAbInitio}).

To evaluate the effect of TM elements
on the conductivity, we have considered an
hypothetical   $\alpha$-Al$_{69.6}$Si$_{13.0}$Cu$_{17.4}$
constructed by
putting Cu atoms in place of Mn atoms in the actual
$\alpha$-Al$_{69.6}$Si$_{13.0}$Mn$_{17.4}$ structure.
Cu atoms have almost
the same number of $sp$ electrons as Mn atoms,
but their $d$ DOS is very small at $\ef$.
Therefore in $\alpha$-Al$_{69.6}$Si$_{13.0}$Cu$_{17.4}$,
the effect of  $sp$(Al)--$d$(TM) hybridization on electronic states
with energy near $\ef$ is very small.
As a result, the pseudogap disappears in total DOS,
and the conductivity is now ballistic (metallic),
$\sigma \simeq \sigma_{\B}$, 
as shown on figure~\ref{Fig_Sig_aAlCuSi}.

\section{Conclusion}

In this article we present the effect of transition-metal atoms on the
physical properties of quasicrystals and related complex phases.
These studies lead  to consider these aluminides as $spd$ electron phases
\cite{PMS05},
where a specific electronic structure governs stability,
magnetism and quantum transport properties.
The principal aspects of this new physics are now understood   
particularly  thanks to  seminal work of 
Prof. T. Fujiwara   
and  subsequent developpements of his ideas.

\bibliographystyle{plain}

\end{document}